# Inverse Scattering Method I.

*Methodological part with an example: Soliton solution of the Sine-Gordon Equation.*

Matej Hudák
*Lab,* Stierova 23, SK-040 23 Košice, Slovak Republic
hudakm@mail.pvt.sk

Jana Tothova
*Lab,* Stierova 23, SK-040 23 Košice, Slovak Republic

Ondrej Hudák*
*Lab,* and Faculty of Aerodynamics, Department of Aviation Technical Studies
Technical University Košice
Rampova 7, SK 040 01 Košice, Slovak Republic
hudako@mail.pvt.sk

March 17, 2018

## Abstract

The aim of this opening paper(I.) is to introduce the Inverse Scattering Method for later studies of some problems in nonlinear dynamics, and describe the kink solution of the Sine-Gordon Equation using the Inverse Scattering Method as an methodological example (the soliton solution is well known). As we discuss in the last Section in the following papers (all in one would be too long paper) we will focus our attention on two regions: the first region of interest is to study solitary traveling wave solutions of pressure equation in bubbly liquids with examination for viscosity and heat transfer. Our second region of interest is to study motions in transport, e.i. in automobiles, trains, aviation and ships, namely nonlinear effects present in it.

## Contents

---

*Corresponding author.

# 1  Introduction.

Russel in 1834 observed a great solitary wave in a channel [1]. A review on soliton and his history may be found in [2]. The author review the history of a soliton, since the mentioned first recorded observation of the great solitary wave by Russell, as a means of developing the mathematical properties of a large class of solvable nonlinear evolution equations. In this class we can find the Korteweg-de Vries, Sine-Gordon, and Nonlinear Schrödinger Equations. Solitary waves, solitons, Bäcklund transformations, conserved quantities and integrable evolution equations as completely integrable Hamiltonian systems are all introduced this way in [2].

In [3] the editors in their book from 2014 besides description of the Sine-Gordon Model, its general background, physical motivations, Inverse Scattering Method and Solitons, included discussions of problems like the Sine-Gordon Equation: From Discrete to Continuum, Soliton Collisions, effects of radiation on Sine-Gordon coherent structures in the continuous and discrete cases, include discussion of experimental results for the Sine-Gordon Equation in arrays of coupled torsion pendula, soliton ratchets in Sine-Gordon-Like Equations, the Sine-Gordon Equation in Josephson-Junction arrays, some selected thoughts old and new on Soliton-Black Hole connections in 2d Dilaton Gravity, dressing with control: using integrability to generate desired solutions to Einstein´s Equations and a planar Skyrme-Like Model.

Thorough survey of physical problems connected with Korteweg de-Vries and modified Korteweg-de Vries Equations, Nonlinear Schrödinger Equation, Sine-Gordon Equation in the form of a single equation, and in the form of systems with coupled equations is discussed by Kivshar and Malomed, [4]. Their discussion includes experimental observations of solitons, the Inverse Scattering Transform and perturbation theory of soliton dynamics in nearly integrable systems.

Let us now mention several examples where the Inverse Scattering Transform and perturbation theory of soliton dynamics in nearly integrable systems is a useful tool.

The Sine-Gordon Equation and the one-dimensional electron gas are related in the Luther-Emery model [5]. Quasi-classical results for this theory indicate the existence of a mass gap for various coupling constants.

Two superconducting metal strips are separated by an insulating barrier which is thin enough $(25 A^o)$ to permit "Josephson" current to tunnel the barrier at superconducting temperatures [6]. Magnetic flux can penetrate along the insulating barrier, and this flux can propagate in the longitudinal $(x)$ direction. There are two possible orientations for the flux. A quantum of flux in one direction is called a "fluxon" and in the other direction an "antifluxon." If an adjustable bias current is present, it will exert a Lorentz force on a fluxon in one direction and on an antifluxon in the other direction. If all dissipative effects and imperfections in the transmission line are neglected, flux propagation is governed by the Sine-Gordon Equation $\Phi_{tt} = \Phi_{xx} - \sin(\Phi)$, where $\Phi$ is magnetic flux measured in units of $\frac{\Phi_0}{2.\pi}$ and $x$ (and $t$) denote suitably normalized space (and time) variables.

An optical three-wave interaction is related to the Sine-Gordon Equation [7]. The authors are discussing solitons in laser physics. They treat the theory of optical self-focusing and filamentation in neutral dielectrics and plasmas.

Here the governing equation is the non-linear Schrödinger Equation or one of its generalizations. They treat the theory of optical self-induced transparency (SIT) at greater length and develop the Maxwell-Bloch (MB), reduced Maxwell-Bloch (RMB), SIT and Sine-Gordon Equations to describe it.

Soliton-like spin waves in $^3He$ are discussed in [8]. Spin waves in $^3HeA$ are governed by the Sine-Gordon Equation. Its kink and breather solutions are elementary nonlinear excitations which generalize the usual concept of spin waves in a linearized theory. Spin waves in $^3HeB$ are governed by a non-integrable "Double Sine-Gordon Equation". Note that the first descriptions of $^3He$ properties did not study nonlinear Sine-Gordon Equation solutions in the form of soliton-like spin waves: a theoretical description of the new phases of liquid $^3He$ is discussed by A.J. Legget in [9] and experimental properties of superfluid $^3He$ are presented in [10]. General gauge invariance and spin waves in the $^3HeB$ phase of superfluid $^3He$ is discussed in [11]. Here the author studied a small oscillation in the Eulerian angles around two equilibrium configurations in the B phase of superfluid $^3He$, which are very likely to be realized in the high-field region and in the low-field region, respectively. Later he studied creation of magnetic solitons in superfluid $^3He$ is studied. The necessary and sufficient conditions for soliton creation in superfluid $^3He$, by turning off an inhomogeneous magnetic field, axe studied theoretically. In the A phase a basic equation is the Sine-Gordon Equation, the initial value problem is solved by making use of the inverse scattering approach. In order to create pair's of solitons, two conditions have to be satisfied: 1, the total impulse (i.e., the spatial integral of the turned-off field) has to be larger than a discrete set of threshold values; and 2, the maximum Zeeman field associated with the turned-off field has to be larger than the threshold value w, which are functions of the total impulse. In the B phase, where no analytical method is available, authors resort to numerical analysis. In spite of significant difference in the dynamic equation, creation of solitons in the B phase can be expressed in terms of two conditions similar to ones in the A phase.

Relationships among Inverse Method, Backlund Transformation and an infinite number of conservation laws are studied in [12]. The main aim of their paper is to clarify the relationships among these properties, i.e., Inverse Method, Backlund transformation and an infinite number of conservation laws. The special interests they pay on the Korteweg-de Vries Equation, the Sine-Gordon Equation and the Modified Korteweg-de Vries Equation.

Light propagation and paired super-radiance in coherent medium studied Yoshimura [13]. Polarization Ri of medium is related to the field $E_0$ by the Maxwell equation. Author [13] shows that the basic problem is to solve nonlinear equation under an arbitrary initial and boundary data. This is the familiar Sine-Gordon Equation for propagation of pulses and SR at the resonant frequency.

The aim of this introductory paper (an opening paper) is to introduce (for our studies of some problems in nonlinear dynamics method) the Inverse Scattering Method [14] - [19], and describe a solution of the Sine-Gordon Equation using this method in the form of the kink (soliton) as an (well known) example. As we discuss in the last Section we then in the following papers focus our attention on two regions: the first region of interest is study of solitary traveling wave solutions of pressure equation in bubbly liquids with examination for viscosity and heat transfer. Our second region of interest is to study motions in transport, e.i in automobiles, trains and ships, namely nonlinear effects.

## 2 Sine-Gordon Equation and Some of its Solutions.

The Sine-Gordon Equation has the form:

$$(\frac{\partial^2}{\partial T^2} - \frac{\partial^2}{\partial X^2})u(X,T) + \sin(u(X,T)) = 0, \tag{1}$$

where X is a space coordinate and T is a time coordinate. For coordinates $(x,t)$ of the "light cone" type:

$$x \equiv \frac{X+T}{2}, t \equiv \frac{X-T}{2} \tag{2}$$

the Sine-Gordon Equation (1) has the form:

$$\frac{\partial^2}{\partial x \partial t} u(x,t) = \sin(u(x,t)). \tag{3}$$

The most known soliton solutions are: kinks, antikinks and breathers. Solution of the kink type of the Sine-Gordon Equation (1) has the form:

$$u(X,T) = 4.\arctan(\exp(\frac{X-U.T}{\sqrt{1-U^2}})). \tag{4}$$

Solution of the breather type of the Sine-Gordon Equation (1) has the form:

$$u(X,T) = 4.\arctan\frac{(\sqrt{\frac{1-\omega^2}{\omega^2}}.\cos(\omega.(T-T_0))}{\cosh(\sqrt{1-\omega^2}.(X-X_0))}. \tag{5}$$

Solution of the N-kink type (N is an integer positive) of the Sine-Gordon Equation (1) has the form:

$$u(X,T) = \cos^{-1}\left(1 + 2.(\frac{\partial^2}{\partial t^2} - \frac{\partial^2}{\partial x^2}).\ln(f(x,t))\right), \tag{6}$$

where:
$$f(x,t) = det(\mathbf{M}),$$

$$M_{k,j} = \frac{\sqrt{a_k a_j}}{a_k + a_j}.\left(\exp\left(K_k.x - \beta_k.t + \gamma_k\right) + (-1)^{k+j}.\exp\left(-K_j.x + \beta_j.t - \gamma_j\right)\right)$$

$$K_j \equiv a_j + \frac{1}{a_j}$$

and where:

$$\beta_j \equiv \frac{1}{a_j} - a_j.$$

Here $\gamma_j$ and $a_j$ are arbitrary constants, and $j = 1, 2, 3, ..., N$. Constants $U(U^2 \leq 1)$ and $\omega(\omega^2 \leq 1)$ are in given intervals.

# 3  Inverse Method of Scattering - Idea.

Our aim is to solve the Sine-Gordon Equation (1) , resp. (3). We will use the fact that it is possible to associate with the equation (1) a system of two linear partial differential equations of the first order, with which a solution of the equation (1) is connected directly as a potential. This is in analogy with the Schrödinger Equation, in which a potential is directly reflecting waves. The Schrödinger Equation is a linear differential equation of the second order. When the potential is known then we can, using known methods, calculate number of localized states and their energies, and coefficients of reflection and transmission, e.i. data which we call scattering data. We are speaking about a Direct Method of Scattering. Inverse Method of Scattering is based on calculations of a potential which is contained in the Schrödinger Equation from the known scattering data. If we take instead the Schrödinger Equation a system of two mentioned above linear equations of the first order and if we modify the apparatus of the Scattering Theory for the Schrödinger Equation to that for these two linear equations, we can solve the Sine-Gordon Equation in the following way.

We know the Sine-Gordon Equation (1) and initial condition. Thus we know the potential in time $t = 0$ in the associated linear problem. Using the Direct Method of Scattering we find the scattering data. They correspond to this potential in the time $t = 0$. We will find equations which determine time development of the scattering data, we will find scattering data for an arbitrary time $t > 0$. Knowing scattering data in the time $t > 0$ we will find using the Inverse Method of Scattering corresponding to this time $t > 0$ the potential, which is as a linear problem directly connected with a solution of the Sine-Gordon Equation, by these steps we thus find a solution of the Sine-Gordon Equation in time $t > 0$. This is the Inverse Method of Scattering idea. Definitions, terms and theorems which we will need in the method will be developed in the text.

# 4  Linear Problem Associated with the Sine-Gordon Equation (1).

Under the term "linear problem associated with the Sine-Gordon Equation (1)" we understand system of two differential equations of the first order, in which there is a solution of the Sine-Gordon Equation appearing in the potential , e.i. in the form $F(u).\mathbf{v}$ , where $F(u)$ is a functional depending on the solution of the Sine-Gordon Equation (1) and $\mathbf{v}$ is a solution of this system, here $\mathbf{v}$ is a two line column vector $\mathbf{v}$ with components $v_1$ (upper component) and $v_2$ (lower component). Explicitly writing we understand under this system the following one:

$$\frac{\partial v_1}{\partial x} + i.\zeta.v_1 = q.v_2, \qquad (7)$$

$$\frac{\partial v_2}{\partial x} - i.\zeta.v_2 = -q.v_1,$$

where

$$q(x,t) \equiv -\frac{1}{2}\frac{\partial}{\partial x}u(x,t)$$

and where $u(x,t)$ is a solution of the Sine-Gordon Equation (1). Here $\zeta$ is a complex number, it is an eigenvalue of the problem. And $v_1, v_2$ are eigenfunctions of the problem (7).

Let us assume that the solution $u(x,t)$ of the Sine-Gordon Equation (1) is the solution corresponding to the boudary conditions:

$$u(x,t) \to 0, \qquad (8)$$

$$\mod (2\pi),$$

for $|x| \to \infty$. Here all derivations of $u(x,t)$ with respect to $x$ are zero for $|x| \to \infty$.

Mathematically saying, (7) is a problem to find eigenvalues $\zeta$ and eigenfunctions $\mathbf{v}$. In physics the problem (7) is a problem of the scattering, this means an analogy with the Schrödinger Equation above and with a physical problem of a particle scattering on a target potential.

Symbolically we can write equations (7) in the following form:

$$\widehat{L}.\mathbf{v} = \zeta \mathbf{v}, \qquad (9)$$

where the operator $\widehat{L}$ has the form of a $2 \times 2$ matrix:

$$\widehat{L} \equiv i. \begin{bmatrix} \frac{\partial}{\partial x} & -q(x,t) \\ -q(x,t) & -\frac{\partial}{\partial x} \end{bmatrix}. \qquad (10)$$

Here

$$\mathbf{v} \equiv \begin{pmatrix} v_1 \\ v_2 \end{pmatrix}. \qquad (11)$$

The system of equations (7) is a system in which time is present as a parameter. Time development of functions $\mathbf{v}$, which solve (7), is chosen in such a way in order to fulfill the following requirement: if $\widehat{A}$ is an operator of a time development of functions $\mathbf{v}(x,t)$ (solutions of the system (7)):

$$\widehat{A}.\mathbf{v} = \frac{\partial}{\partial t}\mathbf{v}, \qquad (12)$$

then this operator have to be so that the operator equality:

$$\widehat{L_t} = [\widehat{A}, \widehat{L}] \qquad (13)$$

is valid for all solutions of the system (7) and gives the Sine-Gordon Equation (1). Here the operator $\widehat{L_t}$ is time derivative of the operator $\widehat{L}$ e.i. $\widehat{L_t} = \frac{\partial \widehat{L}}{\partial t}$. Operator $\widehat{A}$, which fulfill this condition, has the form:

$$\widehat{A} \equiv \frac{i}{4.\zeta}. \begin{bmatrix} \cos(u) & \sin(u) \\ \sin(u) & -\cos(u) \end{bmatrix}. \qquad (14)$$

We will show this fact. Let us take explicit expression of operators $\widehat{L}$ and $\widehat{A}$ from (10) and (14), and substitute them into the equation for time development (13). We obtain that the commutator in (13) has the form:

$$\widehat{L}.\widehat{A} - \widehat{A}.\widehat{L} = -\frac{1}{4.\zeta}. \begin{bmatrix} -u_x.\sin(u) & 2.\sin(u).\frac{\partial}{\partial x} \\ -2.\sin(u).\frac{\partial}{\partial x} & -u_x.\sin(u) \end{bmatrix} = \qquad (15)$$

$$= -\frac{\sin(u)}{2.\zeta} \cdot \begin{bmatrix} \frac{1}{2}.u_x & -\frac{\partial}{\partial x} \\ \frac{\partial}{\partial x} & \frac{1}{2}.u_x \end{bmatrix} =$$

$$= \frac{\sin(u)}{2.\zeta} \cdot \begin{bmatrix} 0 & 1 \\ 1 & 0 \end{bmatrix} .(-i.\widehat{L}).$$

Here $u_x = \frac{\partial u(x,t)}{\partial x}$. The equality (13) has to be valid for solutions of the system of equations (7), or the system of equations (9), then we have:

$$\widehat{L}.\widehat{A} - \widehat{A}.\widehat{L} = \frac{\sin(u)}{2.\zeta} \cdot \begin{bmatrix} 0 & 1 \\ 1 & 0 \end{bmatrix} .(-i.\zeta) = \qquad (16)$$

$$= -\frac{i.\sin(u)}{2} \cdot \begin{bmatrix} 0 & 1 \\ 1 & 0 \end{bmatrix}.$$

The operator $\widehat{L}$ after derivation according to time $t$ has the form:

$$\widehat{L_t} = i. \begin{bmatrix} 0 & \frac{1}{2} \cdot \frac{\partial^2}{\partial x \partial t} u(x,t) \\ \frac{1}{2} \cdot \frac{\partial^2}{\partial x \partial t} u(x,t) & 0 \end{bmatrix}. \qquad (17)$$

Substituting (16) and (17) we obtain:

$$i. \begin{bmatrix} 0 & \frac{1}{2} \cdot \frac{\partial^2}{\partial x \partial t} u(x,t) \\ \frac{1}{2} \cdot \frac{\partial^2}{\partial x \partial t} u(x,t) & 0 \end{bmatrix} = \qquad (18)$$

$$= \frac{i.\sin(u)}{2} \cdot \begin{bmatrix} 0 & 1 \\ 1 & 0 \end{bmatrix}.$$

This is the Sine-Gordon Equation (3).

At the end of this Section we will show an important property of eigenvalues $\zeta$ in the scattering problem (7). The potential $q(x,t)$ is dependent on time, because on time depends $u(x,t)$. From this it may be expected that the eigenvalues $\zeta$ are time dependent. However this is not true. We will show that:

$$\frac{\partial}{\partial t}\zeta = 0. \qquad (19)$$

The proof is straightforward:

$$\frac{\partial}{\partial t}(\widehat{L}.\mathbf{v}) = \widehat{L_t}.\mathbf{v} + \widehat{L}.\frac{\partial \mathbf{v}}{\partial t} = \qquad (20)$$

$$= \frac{\partial}{\partial t}(\zeta.\mathbf{v}) = \frac{\partial \zeta}{\partial t}.\mathbf{v} + \zeta.\frac{\partial \mathbf{v}}{\partial t}.$$

Left side in the equation (20) has the form using equations (12) and (13):

$$\widehat{L_t}.\mathbf{v} + \widehat{L}.\frac{\partial \mathbf{v}}{\partial t} = [\widehat{A}, \widehat{L}].\mathbf{v} + \widehat{L}.\frac{\partial \mathbf{v}}{\partial t} = \widehat{A}.\widehat{L}.\mathbf{v} - \widehat{L}.\widehat{A}.\mathbf{v} + \widehat{L}.\widehat{A}\mathbf{v} = \qquad (21)$$

$$= \widehat{A}.\zeta.\mathbf{v} = \zeta.\frac{\partial \mathbf{v}}{t}.$$

Comparing the right hand side (20) and the result of (21) we obtain (19).

In this Section we have formulated the scattering problem (7) and we have found the time development of its solutions $\mathbf{v}$, the equation (13) gives the Sine-Gordon Equation. Moreover it was shown that the eigenvalue $\zeta$ of this problem does not depend on time. It will be show that this property is connected with soliton properties of the solutions of the problem.

# 5 Direct Method of the Scattering for the Associated Problem.

Under a Direct Method of Scattering we understand to find scattering data knowing a potential in the Schrödinger Equation. Scattering data are: number and energies of the localized (bounded) states, coefficients of reflection and transmission, and an asymptotic behavior of the wave functions. For the system of differential equations (7) we will define all analogy terms, of Scattering Theory, in the same way as they appear for the Schrödinger Equation.

Let us define a system of equations more general than the system (7):

$$\frac{\partial v_1}{\partial x} + i.\zeta.v_1 = q.v_2, \tag{22}$$

$$\frac{\partial v_2}{\partial x} - i.\zeta.v_2 = -q^*.v_1,$$

where $q(x,t)$ potential may be in general complex, here $^*$ means a complex conjugation. It is immediately possible to demonstrate using (22) the following claimings:

I. Let $\mathbf{v}(\mathbf{w})$ be the solution of equations (22) with the corresponding eigenvalues $\zeta_1(\zeta_2)$. Then the following relation is valid:

$$\frac{\partial}{\partial x}(v_1.w_2 - w_1.v_2) + i.(\zeta_1 - \zeta_2).(v_1.w_2 + w_1.v_2) = 0, \tag{23}$$

where $v_i(w_i)$ with $i = 1, 2$ are components of $\mathbf{v}(\mathbf{w})$.

II. Let $\mathbf{v}$ be the solution of the equations (22) for the eigenvalue $\zeta = \xi_1 + i.\eta_1$ (here $\xi_1, \eta_1$ are real numbers.) Then the following relation is valid defining operation:

$$\overline{\mathbf{v}} \equiv \begin{pmatrix} v_2^* \\ -v_1^* \end{pmatrix} \tag{24}$$

and

$$\overline{\overline{\mathbf{v}}} = -\mathbf{v}$$

is a solution of the system of equations (22) with the following properties: a. the corresponding eigenvalue is $\zeta = \xi_1 - i.\eta_1$, and b. if $\eta_1 = 0$ then $\mathbf{v}$ and $\overline{\mathbf{v}}$ are two linearly independent solutions of (22).

From all possible solutions of the equations (22) we will choose special cases, which will be called in the following Jost functions. These solutions of (22) are fulfilling:

- They solve the equations (22),
- For $\zeta = \xi$ real numbers the asymptotic behavior of solution functions is:

$$\varphi \longrightarrow \begin{pmatrix} 1 \\ 0 \end{pmatrix} . \exp(-i.\xi.x) \tag{25}$$

for $x \longrightarrow -\infty$ and $\varphi \equiv \varphi(x, \xi)$ and where:

$$\psi \longrightarrow \begin{pmatrix} 0 \\ 1 \end{pmatrix} . \exp(+i.\xi.x) \tag{26}$$

for $x \longrightarrow +\infty$ and $\psi \equiv \psi(x,\xi)$.

Functions $\varphi(x,\xi)$ and $\psi(x,\xi)$ may be analytically extended in the eigenvalue parameter $\zeta = \xi$ in upper complex half-plane. Denoting Jost functions using $\varphi(x,\xi)$ and $\psi(x,\xi)$ means in the following their asymptotic behavior in $x \longrightarrow -\infty$ and $x \longrightarrow +\infty$.

From the II. statement we know that for real eigenvalues $\zeta = \xi$ the solutions $\mathbf{v}$ and $\overline{\mathbf{v}}$ are linearly independent solutions. While (22) is the system of linear differential equations of the first order (equivalent to one linear equation of the second order), then the full system of solutions of (22) will consists from arbitrary solutions of (22), which are linearly independent. The Jost function $\varphi$ ($\psi$) is the solution of equations (22), from its definition the full system of solutions is $\varphi$ and $\overline{\varphi}$, $\psi$ and $\overline{\psi}$. It means that if we take arbitrary solution of (22), then it is possible to express this solution as a linear combination of Jost functions $\varphi$ and $\overline{\varphi}$, $\psi$ and $\overline{\psi}$. Let as take the Jost function $\varphi$. It is a solution of equations (22) according to its definition, it is possible to express it as a linear combination of functions $\psi$ and $\overline{\psi}$:

$$\varphi = a(\xi).\overline{\psi}(x,\xi) + b(\xi).\psi(x,\xi), \qquad (27)$$

where $a(\xi)$ and $b(\xi)$ are coefficients of the linear combination, and the argument $\xi$ is written here because of their possible dependence on the eigenvalue $\xi$. These coefficients $a(\xi)$ and $b(\xi)$ will play an important role in following parts of the paper, we will show some of their properties.

III. For coefficients $a$ and $b$ defined in (27) the following relation holds:

$$\mid a(\xi) \mid^2 + \mid b(\xi) \mid^2 = 1. \qquad (28)$$

The proof is the following. From (27) we have for $\zeta = \xi$ that $\varphi = a(\xi).\overline{\psi} + b(\xi).\psi$. From this and from (24) we find that $\overline{\varphi} = -a^*(\xi).\psi + b^*(\xi).\overline{\psi}$. and $\zeta^* = \zeta = \xi$. From I. it follows that for $\zeta_1 = \zeta_2 = \xi$ and for two solutions $\varphi$ and $\overline{\varphi}$ of (22) we have $\frac{\partial}{\partial x}(\varphi_1.\overline{\varphi}_2 - \varphi_2.\overline{\varphi}_1) = 0$. From that we find $(\varphi_1.\overline{\varphi}_2 - \varphi_2.\overline{\varphi}_1) = const.$. The value of a constant $const.$ may be calculated from the asymptotic behavior of Jost functions $\varphi$ and $\overline{\varphi}$ for the limit $x \longrightarrow -\infty$:

$$\begin{pmatrix} \varphi_1 \sim \exp(-i.\xi.x) & \overline{\varphi}_1 = \varphi_2^* \sim 0 \\ \varphi_2 \sim 0 & \overline{\varphi}_2 = -\varphi_1^* \sim -\exp(+i.\xi.x) \end{pmatrix}, \qquad (29)$$

which gives $const. = -1$. On the other side we know that (27) holds, and that a similar relation holds for $\overline{\varphi}$, see beginning of the proof, and this leads to:

$$-1 = (\varphi_1.\overline{\varphi}_2 - \varphi_2.\overline{\varphi}_1) = -(a.a^* + b.b^*).(\psi_1.\psi_1^* + \psi_2.\psi_2^*).$$

From the asymptotic behavior of Jost functions $\psi$ and $\overline{\psi}$ for $x \longrightarrow +\infty$ we obtain , see (24) and (26):

$$\begin{pmatrix} \psi_1 \sim 0 & \overline{\psi}_1 \sim \exp(-i.\xi.x) \\ \psi_2 \sim \exp(+i.\xi.x) & \overline{\psi}_2 \sim 0 \end{pmatrix}, \qquad (30)$$

which gives $\psi_1.\psi_1^* + \psi_2.\psi_2^* \sim +1$. That is the reason that $-1 = -(\mid a(\xi) \mid^2 + \mid b(\xi) \mid^2).(+1)$.

IV. If $\varphi$ and $\psi$ are Jost functions from (25) and (26) then for $\zeta = \xi$ we have:

$$a(\xi) = \varphi_1(x,\xi).\psi_2(x,\xi) - \varphi_2(x,\xi).\psi_1(x,\xi). \tag{31}$$

Proof: From I. we know that $\varphi_1(x,\xi).\psi_2(x,\xi) + \varphi_2(x,\xi).\psi_1(x,\xi) = const.$ for two Jost functions $\varphi(x,\xi)$ and $\psi(x,\xi)$ for $\zeta = \xi$. The value of the constant $const.$ we calculate again from the asymptotic behavior of Jost functions for $x \longrightarrow +\infty$, we will use (27):

$$const. = (a(\xi).\overline{\psi_1} + b(\xi).\psi_1).\psi_2 - (a(\xi).\overline{\psi_2} + b(\xi).\psi_2).\psi_1 = \tag{32}$$

$$= a.(\psi_2.\psi_2^* + \psi_1^*.\psi_1) = a,$$

see the proof of III.

V. Coefficient $a$ in (27) is dependent on $\zeta = \xi$ and it is possible analytically extend it into upper complex halfplane in the complex plane $\zeta = \xi + i.\eta$. Proof: From the definition of Jost functions we know that $\varphi$ and $\psi$ functions may be analytically extend into upper complex halfplane $\eta > 0$. Using (31) it is clear that V. is valid.

VI. It is valid:

$$a(\zeta) \longrightarrow 1 \tag{33}$$

for

$$|\zeta| \longrightarrow +\infty$$

$$Im\zeta \geq 0.$$

Proof will be done later.

*Discrete values of the problem* (1). These are eigenvalues $\zeta_j$ where $j = 1, 2, 3, ..., N$ with the following properties:

a. $Im\zeta_j > 0$,

b. $a(\zeta_j) = 0$, here $a$ is a function of the complex argument $\zeta$ with $Im\zeta \geq 0$.

We say that in the points $\zeta = \zeta_j$, where $j = 1, 2, 3, ..., N$, the problem of scattering (1) has bounded states. By this we understand the fact that for $Im\zeta \geq 0$ Jost functions for $|x| \longrightarrow +\infty$ have exponentially decaying dependence. The following statement holds:

VII. If $\zeta_j$ is discrete eigenvalue of the problem (1) then there exist a constant $c_j$ such that

$$\varphi(x,\zeta_j) = c_j\psi(x,\zeta_j). \tag{34}$$

Proof: Here the functions $\varphi(x,\zeta_j)$ and $\psi(x,\zeta_j)$ are Jost functions (25) and (26) respectively, they are defined by their asymptotic behavior for large $|x|$ for $\zeta = \xi$ real and analyticaly continued to upper complex halfplane $\eta \geq 0$, where $\zeta = \xi + i.\eta$. Because for the discrete value $\zeta_j$ the equation $a(\zeta_j) = 0$ is valid, and because the relation (31) can be analyticaly extended from $\zeta = \xi$ to $\zeta = \xi + i\eta$ with $\eta \geq 0$, then for $\zeta = \zeta_j$ the following relation is valid:

$$\varphi_1(x,\zeta_j).\psi_2(x,\zeta_j) = \varphi_2(x,\zeta_j).\psi_1(x,\zeta_j) \tag{35}$$

from which it follows (34).

If we confine ourselves in the problem (22) on the real potential, as it is the case in the Sine-Gordon Equation, then there exist additional statements about Jost functions and the coefficient $a$:

VIII. Nulls of the coefficient $a(\zeta)$ are lying on the imaginary axis $\zeta_j = i.\eta_j$, where $\eta > 0$, or in the upper halfplane of the complex plane $\zeta$ in complex conjugated points $\zeta_j$ and $\zeta_j^*$. Here $j = 1, 2, 3, ..., N$. Proof: Directly from the (22) it follows that for the real potential $q$ we have $\varphi(x, -\xi) = \varphi^*(x, \xi)$ and $\psi(x, -\xi) = \psi^*(x, \xi)$. From this it follows, using (31), that:

$$a(\xi) = a^*(-\xi). \tag{36}$$

This equation after analytically extending into the upper complex halfplane $\eta > 0$ gives:

$$a(\zeta) = a^*(-\zeta^*) \equiv a^*(-\xi + i\eta) = a^*(\xi + i\eta).$$

If $\zeta_j$ is a discrete eigenvalue of the problem (22) then $a(\zeta_j = 0)$ and from (36) we have $a^*(-\zeta_j^* = 0)$. Simultaneously from $a(\zeta_j = 0)$ it follows that $a^*(\zeta_j = 0)$. It means that either we have:

$$-\zeta_j^* = -(\xi_j - i.\eta_j) = \xi_j + i.\eta_j = \zeta_j,$$

and from this $\xi_j = 0$ and $\zeta_j = i.\eta_j$, or zeros are in $\zeta_j$ $(a(\zeta_j) = 0)$ and in $-\zeta_j^*$ because $(a^*(-\zeta^*))_j = 0)$ from which we obtain $(a(-\zeta_j^*))$.

Let us now describe "scattering data" and let us find connection between the problem (22) and terms of the scattering problem. The system of equations (22) is assumed here to be analogous with the Schrödinger Equation. If we know the potential q then we may search solutions of the problem. Jost functions $\varphi$, $\overline{\varphi}$ and $\psi$, $\overline{\psi}$ are the full system of solutions of equations (22), thus it is enough to search only for these functions. If we find this system solving the equations (22) with boundary conditions (25) and (26) then we can using (27) find the coefficients $a$ and $b$. After analytical continuation of the coefficient $a$ into upper $\zeta$ half complex plane we find bounded states, characterized by $c_j$ and $\zeta_j$ where $j = 1, 2, 3, ..., N$ and corresponding Jost functions using (36). We say that we have found scattering data of the problem (22):

the scattering data of the problem (22) are $a(\xi)$, $b(\xi)$, $\zeta_j$, $c_j$, where $j = 1, 2, 3, ..., N$ and where $|\xi| < \infty$.

Why we say scattering data? Because we use the following interpretation of the equation (27):

$$\frac{1}{a}.\varphi = \overline{\psi} + \frac{b}{a}.\psi, \tag{37}$$

here $\zeta = \xi$. The Jost function $\overline{\psi}$ describes particles falling in the region where our potential q is nonzero. Part of these particles is transmissing the region and part is reflecting back (this is 1-dimensional case, scattering into different angles is not present here). Transmitting part of particles is described by the wave function $\frac{1}{a}\psi$ and reflecting part is described by $\frac{b}{a}\psi$. The coefficient $|a|^2$ plays the role of the coefficient of transmission and the coefficient $|b|^2$ the role of the coefficient of reflection.

Direct problem of the Theory of Scattering is in the case (22) a problem to calculate from the known potential $q$ scattering data. Inverse problem of the Theory of Scattering is a problem to calculate from the known scattering data the potential $q$ (to reconstruct the potential).

let us note at the end of this Section that the problem (22) depends on the parameter $t$, e.i. on time. We have solved this problem (22) for fixed $t$. If time in the potential $q$ is changing scattering data are changing too. How these scattering data are changing with time $t$ will be discussed in the following Section.

## 6 Time Development of the Scattering Data.

Let us discuss time development of the scattering data.

- The case $\xi$ is real. Time development of scattering data will be found from the equation for time development of eigenfunctions in (22). We will use the equation (1) and the asymptotic behavior of Jost functions $\varphi$ and $\psi$. Let us denote by $v$:

$$v \equiv \overline{\psi} + \frac{b}{a}\psi. \tag{38}$$

For $x \longrightarrow +\infty$ from the definition of the Jost function $\psi$:

$$v \approx \begin{pmatrix} 1 \\ 0 \end{pmatrix}.\exp(-i.\xi.x).f(x,t) + \frac{b}{a}\begin{pmatrix} 0 \\ 1 \end{pmatrix}.\exp(+i.\xi.x).f(x,t) \equiv \tag{39}$$

$$\equiv \begin{pmatrix} v_1 \\ v_2 \end{pmatrix}$$

where the function $f(x,t)$ has asymptotic behavior

$$\lim_{|x|\longrightarrow +\infty} (f(x,t)) = +1$$

for every $t \geq 0$. In this asymptotic region $\mid x \mid \longrightarrow +\infty$ have equations (12) the form:

$$\frac{\partial}{\partial t}v_1 = \frac{i}{4.\zeta}v_1, \tag{40}$$

$$\frac{\partial}{\partial t}v_2 = -\frac{i}{4.\zeta}v_2.$$

because (potential $q = 0$ and) $u(x,t) \longrightarrow 0$ with $mod(2.\pi)$ in the matrix $\widehat{A}$. After substituting (39) into (40) we obtain:

$$\exp(-i.\xi.x)\frac{\partial f}{\partial t} = \frac{i}{4.\zeta}.\exp(-i.\xi.x).f \tag{41}$$

and from this we obtain $\frac{\partial f}{\partial t} = \frac{i}{4.\zeta}.f$ and

$$\exp(+i.\xi.x)(\frac{\partial f}{\partial t}.\frac{b}{a} + f.\frac{\partial}{\partial t}(\frac{b}{a})) = -\frac{i}{4.\zeta}\frac{b}{a}.f.\exp(+i.\xi.x)$$

. From this and from the equation (41) we find:

$$\frac{\partial}{\partial t}(\frac{b}{a}) = -\frac{2.i}{4.\zeta}.(\frac{b}{a}) = -\frac{i}{2.\zeta}.(\frac{b}{a}). \tag{42}$$

It is possible to find time development of the coefficient $a$ using statement IV. from the previous Section:

$$a(\xi, t) = \varphi_1(x, t, \xi).\psi_2(x, t, \xi) - \varphi_2(x, t, \xi).\psi_1(x, t, \xi) \tag{43}$$

and using asymptotic behavior of the Sine-Gordon Equation solutions for $|x| \longrightarrow +\infty$ which is $u(x, t) \longrightarrow 0$ modulo mod $(2\pi)$ in the equations for time development of Jost functions $\varphi$ and $\psi$, e.i.

$$\frac{\partial \varphi_1}{\partial t} = \frac{i}{4.\zeta}.\varphi_1, \tag{44}$$

$$\frac{\partial \varphi_2}{\partial t} = -\frac{i}{4.\zeta}.\varphi_2,$$

$$\frac{\partial \psi_1}{\partial t} = \frac{i}{4.\zeta}.\psi_1,$$

$$\frac{\partial \psi_2}{\partial t} = -\frac{i}{4.\zeta}.\psi_2,$$

for $|x| \longrightarrow +\infty$.

If we derive with respect to time the left and the right side of the equation (43) then we obtain for $x \longrightarrow +\infty$ (or for $x \longrightarrow -\infty$, it is not important which case we take in our calculations: it is important that $u \longrightarrow 0$ with mod $(2\pi)$ ):

$$\frac{\partial a(\xi, t)}{\partial t} = \frac{\partial \varphi_1}{\partial t}.\psi_2 + \varphi_1.\frac{\partial \psi_2}{\partial t} - \tag{45}$$

$$-\frac{\partial \varphi_2}{\partial t}.\psi_1 - \varphi_2.\frac{\partial \psi_1}{\partial t} =$$

$$= \frac{i}{4.\zeta}(\varphi_1.\psi_2 - \varphi_1.\psi_2 - (-\varphi_2).\psi_1) - \varphi_2.\psi_1 = 0.$$

It means that:

$$\frac{\partial a(\xi, t)}{\partial t} = 0. \tag{46}$$

From this equation (46) we obtain

$$a(\xi, t) = a(\xi, 0).$$

It follows from (42) and (46) that:

$$\frac{\partial b}{\partial t} = -\frac{i}{2.\zeta}.b, \tag{47}$$

which gives $b(\xi, t) = \exp(-\frac{i.t}{2.\xi}).b(\xi, 0)$. Equations (46) and (47) give time evolution of coefficients $a(\xi, t)$ and $b(\xi, t)$ for eigenvalue $\zeta = \xi$ real.

- The case $\zeta = \zeta_j$ with $Im\zeta_j > 0$. Coefficients $c_j$ are analytical continuation of the coefficient $b(\xi)$ into upper half-plane with $\eta > 0$, and namely into the point $\zeta = \zeta_j$, where $a = 0$. It means that:

$$c_j = b(\zeta_j, t) = \exp(-\frac{i.t}{2.\zeta_j}).b(\zeta_j, 0) = \exp(-\frac{i.t}{2.\zeta_j}).c_j(t=0). \qquad (48)$$

Let us note that deriving time dependence of the scattering data of the problem (22) we used the fact that eigenvalues of this problem are independent on time $t$, this fact is not clear from the first sight because the potential $q$ is time dependent in general.

## 7 Inverse Method of the Scattering Theory for the Problem (1).

We will study the following problem in this Section. Knowing the scattering data $a(\xi, t)$, $b(\xi, t)$ for $|\xi| < \infty$, and $c_j(t)$ and $\xi_j$ for $j = 1, 2, 3, ..., N$ we would like to know which potential $q(x, t)$ in the system (22) corresponds to these scattering data.

Let us note that values of scattering data we can calculate by the Direct Method of the Theory of Scattering from a given initial condition of the Sine-Gordon Equation in time $t = 0$: $u(x, t = 0)$. This function determines the potential in the problem (22) and corresponding scattering data will be determined in time $t = 0$. In the previous Section we have found time development of the scattering data for $t > 0$, in this Section we will find from this scattering data the potential $q(x, t)$. Thus we will be able to find $u(x, t)$ for $t > 0$ which is a solution of the Sine-Gordon Equation.

There are following steps in the Inverse Method of the Theory of Scattering. First we will show, how it is possible to reconstruct an analytic function if we know its residua in simple poles and its jump through the real axis. Then we find a system of equations of the Inverse Method for calculation of this jump and for calculation of the values of this function in its poles for a given parameters of the scattering.

Let us define the function $\Phi(\zeta)$ of the complex variable $\zeta$ in this way:

$$\Phi(\zeta) \equiv \Phi(\zeta, x) = \frac{1}{a(\zeta)}.\varphi(x, \zeta).\exp(+i.\zeta.x) \qquad (49)$$

for $Im\zeta > 0$, and

$$\Phi(\zeta) \equiv \Phi(\zeta, x) = \overline{\psi(x, \zeta)}.\exp(+i.\zeta.x)$$

for $Im\zeta < 0$. On the real axis $Im\zeta = 0$, $\zeta = \xi$ we have:

$$\Phi(\xi) = \frac{1}{2}.[\lim_{\epsilon \longrightarrow 0_+}(\Phi(\xi + i.\epsilon)) + \Phi(\xi - i.\epsilon))]. \qquad (50)$$

As a jump of the function $\Phi(\zeta)$ going through the real axis $\zeta = \xi$ we understand the function $\Theta(\xi)$ defined in the following way:

$$\Theta(\xi) \equiv [\lim_{\epsilon \longrightarrow 0_+}(\Phi(\xi + i.\epsilon)) - \Phi(\xi - i.\epsilon))]. \qquad (51)$$

In (49), (50) and (51) there is everywhere present variable $x$ as a fixed parameter.

In the following we will assume that zeros of $\zeta_1, \zeta_2, \zeta_3, ..., \zeta_N$ of the function $a(\zeta)$ are simple. The following statement holds:

- IX Function $\Phi(\zeta)$ analytical piecewise can be expressed through the jump of this function, and through residua in poles $\zeta_j$ where $j = 1, 2, 3, ..., N$ for every $\zeta$ as:

$$\Phi(\zeta) = \sum_{n=1}^{n=N} \frac{\exp(+i.\zeta_n.x)}{\zeta - \zeta_n} \cdot \frac{\varphi(x, \zeta_n)}{\frac{da(\zeta=\zeta_n)}{d\zeta}} + \qquad (52)$$

$$+ \frac{1}{2.\pi.i} \int_{-\infty}^{+\infty} \frac{\Phi(\xi)}{\xi - \zeta} + \begin{pmatrix} 1 \\ 0 \end{pmatrix}.$$

The proof of IX is straightforward.

Let us introduce the following denotations:

$$\Phi^{(1)}(\zeta, x) \equiv \begin{pmatrix} 1 \\ 0 \end{pmatrix} + \sum_{n=1}^{n=N} \frac{\exp(+i.\zeta_n.x)}{\zeta - \zeta_n} \cdot \widetilde{c}_n \cdot \psi(x, \zeta_n), \qquad (53)$$

$$\Phi^{(2)}(\zeta, x) \equiv \frac{1}{2.\pi.i} \cdot \int_{-\infty}^{+\infty} \frac{\Theta(\xi)}{\xi - \zeta} .d\xi,$$

$$\widetilde{c}_n \equiv \frac{c_n}{\frac{da(\zeta)}{d\zeta}|_{\zeta=\zeta_n}}.$$

In (53) we used the fact that for $\zeta = \zeta_j$ we have $\phi(x, \zeta_j) = c_j.\psi(x, \zeta_j)$, here $j = 1, 2, 3, ..., N$., see Statement $VII$. For simplicity the tilde over $c_n$ in $\widetilde{c}_n$ is omitted in the following.

Now we can write (52) in the form:

$$\Phi(\zeta, x) = \Phi^{(1)}(\zeta, x) + \Phi^{(2)}(\zeta, x). \qquad (54)$$

Let us calculate now the step (jump) $\Theta(\xi)$:

$$\Theta(\xi) \equiv \lim_{\epsilon \to 0_+} [\Phi^{(1)}(\zeta + i.\epsilon) - \Phi^{(2)}(\zeta - i.\epsilon)] = \qquad (55)$$

$$= \frac{\exp(+i.\xi.x).\varphi(x, \xi)}{a(\xi)} - \overline{\psi(x, \xi)}. \exp(+i.\xi.x) =$$

$$= \exp(+i.\xi.x). \frac{b(\xi)}{a(\xi)} .\psi(x, \xi),$$

where we used the definition of the function $\Phi$ (48) and the relation between the Jost functions $\phi$, $\psi$ and $\overline{\psi}$.

The equation (55) shows that if $b = 0$ (in this case we speak about reflexless potential q, see the Section about the Direct Method of the Scattering Theory for the associated problem), then $\Theta(\xi) = 0$.

Statement $IX$ enabled to find the function $\Phi(\zeta, x)$ (and thus Jost functions $\phi$ and $\psi$, see (49)) through $\psi(x, \zeta_n)$ and $\Theta(xi)$ and through the scattering data. Let us show how it is possible to calculate functions $\psi(x, \zeta_n)$ and $\Theta(\xi)$ for given scattering data.

## 7.1 The System of Equations of the Inverse Method of Theory of Scattering
for Calculation of Functions $\psi(x, \zeta_n)$ and $\Theta(\xi)$ for Given Scattering Data.

Our aim is to find equations for the calculation of functions $\psi(x, \zeta_n)$, where $n = 1, 2, 3, ...N$, and $\Theta(\xi)$ which enable to reconstruct the function $\Phi(\zeta)$. let us discuss two cases, 1. where $\zeta = \xi$, and 2. where $\zeta = \zeta_j$ with $j = 1, 2, 3, ...N$.

- 1. where $\zeta = \xi$ On the real axis we defined the function $\Phi(\xi, x)$ in (50). Using (52) we obtain:

$$\Phi(\xi) = \Phi^{(1)} + \frac{1}{2}(J\Theta)(\xi), \tag{56}$$

where $J$ is the Hilbert transformation:

$$(J\Theta)(\xi) \equiv \frac{1}{\pi.i} \int_{-\infty}^{+\infty} \frac{\Theta(\xi')}{\xi' - \xi} d\xi'. \tag{57}$$

Note that $(J\Theta)^* = -(J\Theta^*)$. Besides the equation (52) we can use the equation (50) and the definition of the function $\Phi(\zeta)$ in (49). This leads using (55) to:

$$\Phi(\xi) = \frac{1}{2}.(\frac{\varphi(x,\xi)}{a(\xi)}.\exp(+i.\xi.x) + \overline{\psi}(x,\xi).\exp(+i.\xi.x)) = \tag{58}$$

$$= \frac{1}{2}(\Theta(\xi) + 2.\overline{\psi}(x,\xi).\exp(+i.\xi.x)).$$

We obtain from (56) and (58):

$$\overline{\psi}(x,\xi).\exp(+i.\xi.x) = -\frac{1}{2}(1-J)\Theta(\xi,x) + \Phi^{(1)}(\xi,x). \tag{59}$$

This relation we can write in components, the second component will be complex conjugated:

$$\psi_2^*(x,\xi).\exp(+i.\xi.x) + \frac{1}{2}(1-J)\Theta_1(\xi,x) = \Phi_1^{(1)}(\xi,x), \tag{60}$$

$$-\psi_1(x,\xi).\exp(-i.\xi.x) + \frac{1}{2}(1+J)\Theta_2^*(\xi,x) = \Phi_2^{(1)}(\xi,x),$$

where the lover index denotes the component. We used here properties of the Hilbert transformation, see (57).

In (60) we multiply the raw by the function $c^*(x,\xi)$ and the second by the function $c(x,\xi)$, where:

$$c(x,\xi) \equiv \frac{b(\xi)}{a(\xi)}.\exp(+2.i.\xi.x) \tag{61}$$

and the resulting system of two equations is (see (57)):

$$\Theta_1(\xi,x) - c(x,\xi).\frac{1+J}{2}\Theta_2^*(\xi,x) = \tag{62}$$

$$= -c(x,\xi).\sum_{n=1}^{n=N}\frac{\exp(-i.\zeta_n^*.x)}{\xi-\zeta_n^*}.c_n^*.\psi_2^*(x,\zeta_n)$$

and

$$\Theta_2^*(\xi,x) + c^*(x,\xi).\frac{1-J}{2}\Theta_1(\xi,x) = \qquad (63)$$

$$= c^*(x,\xi) + c^*(x,\xi).\sum_{n=1}^{n=N}\frac{\exp(+i.\zeta_n.x)}{\xi-\zeta_n^*}.c_n.\psi_1^*(x,\zeta_n).$$

- 2. where $\zeta = \zeta_j$ with $j = 1,2,3,...N$.

  From (49) and (52) for discrete eigenvalues we obtain:

$$\psi_1(x,\zeta_j).\exp(-i.\zeta_j.x) + \sum_{n=1}^{n=N}.\frac{\exp(-i.\zeta_n^*.x)}{\zeta_j-\zeta_n^*}.c_n^*.\psi_2^*(x,\zeta_n) = \qquad (64)$$

$$= \frac{1}{2.\pi.i}\int_{-\infty}^{+\infty}\frac{\Theta_2^*(\xi,x)}{\xi-\zeta_j}d\xi$$

and

$$\psi_2^*(x,\zeta_j).\exp(+i.\zeta_j^*.x) - \sum_{n=1}^{n=N}.\frac{\exp(+i.\zeta_n.x)}{\zeta_j^*-\zeta_n}.c_n.\psi_1(x,\zeta_n) = \qquad (65)$$

$$= 1 + \frac{1}{2.\pi.i}\int_{-\infty}^{+\infty}\frac{\Theta_1(\xi,x)}{\xi-\zeta_j^*}d\xi.$$

The equations (62), (63), (64) and (65) are representing the system of equations for the Inverse Method of the Scattering Theory. The system consists of $2.N+2$ equations for $2.N+2$ unknown: $2.N$ constants $\psi_{1,2}(x,\zeta_j)$ for $j = 1,2,3,...,N$ and two functions $\Theta_{1,2}(\xi,x)$ for $|\xi|<+\infty$, here $x$ is fixed in all steps of this method. It is assumed that parameters of scattering $a(\xi)$, $b(\xi)$, $\zeta_1,\zeta_2,\zeta_3,...,\zeta_N$ and $c_1,c_2,c_3,...,c_N$ are known. Similarly as $x$ the time variable $t$ is fixed, dependence of parameters of scattering see in the Section on Time Development of the Scattering Data, and unknown functions $\phi$, $\psi$ and $\Theta$.

# 8 Calculation of the Potential $q(x,t)$ and Solution of the Sine-Gordon Equation, Marchenko Equation.

In the previous Section we have shown how from the scattering data it is possible to reconstruct "wave" function $\Phi(\zeta,x)$. Knowing this function and using the asymptotic behavior of the Jost function $\psi(x,\zeta)$ for $\zeta = R.\exp(i.\varphi)$ and from the definition of the function $\Phi(\zeta,x)$, and using the asymptotic expansion for $R \longrightarrow +\infty$, where $\zeta = R.\exp(i.\varphi)$ in (52), we obtain for $\varphi \in <\pi, 2.\pi>$:

$$\begin{pmatrix}\psi_2(x,\zeta)\\-\psi_1(x,\zeta)\end{pmatrix}.\exp(-i.\zeta.x) \approx \begin{pmatrix}1\\0\end{pmatrix} + \qquad (66)$$

$$+\frac{1}{R.\exp(i.\varphi)}.[\sum_{n=1}^{n=N} c_n^*.\exp(-i.\zeta^*.x).\psi^*(x,\zeta_n)+$$

$$+\frac{1}{2.\pi.i}\int \Theta^*(\xi,x)d\xi]+O(\frac{1}{R^2}).$$

From this equations we obtain:

$$\begin{pmatrix} \psi_1(x,\zeta) \\ \psi_2(x,\zeta) \end{pmatrix}.\exp(-i.\zeta.x)\approx \begin{pmatrix} 0 \\ 1 \end{pmatrix}+ \qquad (67)$$

$$+\frac{1}{2.i.R.\exp(i.\varphi)}.\begin{pmatrix} q(x,t) \\ \int_x^{+\infty} \mid q(x',t)\mid^2 dx' \end{pmatrix}+O(\frac{1}{R^2}).$$

From comparison of both expansions above we obtain:

$$q(x,t)=-2.i.\sum_{n=1}^{n=N} c_n^*.\exp(-i.\zeta_n^*.x).\psi_2^*(x,\zeta_n)-\frac{1}{\pi}.\int_{-\infty}^{+\infty}\Theta_2^*(\xi,x)d\xi \qquad (68)$$

and

$$\int_x^{+\infty}\mid q(x',t)\mid^2 dx' = -2.i.\sum_{n=1}^{n=N} c_n.\exp(i.\zeta_n.x).\psi_1(x,\zeta_n)+\frac{1}{\pi}\int_{-\infty}^{+\infty}\Theta_1(\xi,x)d\xi. \qquad (69)$$

When we know the potential $q$, then we may to find the solution $u(x,t)$ of the Sine-Gordon Equation in any time $t>0$:

$$q(x,t)=-\frac{1}{2}\frac{\partial}{\partial x}u(x,t). \qquad (70)$$

## 8.1 Marchenko Equation.

Sometimes introduced is another form of equations for calculation of the potential $q(x,t)$, so called Marchenko Equation. Let us define functions $F(x)$ and $K(x,s)$:

$$F(x)\equiv \frac{1}{2.\pi}\int_{-\infty}^{+\infty}\frac{b(\xi)}{a(\xi)}.\exp(+i.\xi.x)d\xi+\sum_{n=1}^{n=N} c_n.\exp(+i.\zeta_n.x) \qquad (71)$$

and

$$\psi(x,\zeta)=\exp(+i.\zeta.x)+\int_x^{+\infty} K(x,s),\exp(+i.\zeta.s)ds, \qquad (72)$$

where $Im\zeta \geqslant 0$. Equations (62), (63), (64) and (65) after Fourier transform will have the form, using definitions of the function $F(x)$ and the kernel $K(x,s)$,:

$$K_1(x,y)=F^*(x+y)+\int_x^{\infty} K_2(x,s).F^*(s+y)ds, \qquad (73)$$

$$K_2(x,y)=-\int_x^{\infty} K_1(x,s)F(s+y).ds.$$

Equations (68) and (69) have the form:

$$q(x,t)=-2.K_1(x,x), \qquad (74)$$

$$\int_{x}^{+\infty} |q(x',t)|^2 \, dx' = -2.K_2(x,x).$$

The equation (72) may be written in the form (this is the Marchenko Equation for $K_1(x,x)$):

$$K_1(x,x) = F^*(x+y) - \int_x^{+\infty} \int_x^{+\infty} F^*(y+z).F(s+z).K_1(x,s) dz ds. \quad (75)$$

## 9 Determination of the Scattering data from the Initial Conditions in $(X,T)$ Coordinates.

We have worked up till now with the Sine-Gordon Equation in coordinates of the light cone $(x,t)$ in which the equation has the form:

$$\frac{\partial^2}{\partial x \partial y} u(x,t) = \sin(u(x,t)). \quad (76)$$

We are interested in the case of $(X,T)$ coordinates, see the Section 1. In these coordinates the Sine-Gordon Equation has the form:

$$(\frac{\partial^2}{\partial T^2} - \frac{\partial^2}{\partial X^2}) u(X,T) + \sin(u(X,T)) = 0, \quad (77)$$

knowing the initial condition:

$$u(X, T=0). \quad (78)$$

To solve the Sine-Gordon Equation in the form (77) we would like to use equations found for coordinates $(x,t)$. We will do it in the following way. Using the Direct Method of the Theory of Scattering for $T=0$ we will find scattering data $(A, B, C_j)$ in analogy to data $(a(\xi), B(\xi), c_j)$, where $j = 1,2,3,...,N$. Using relation, which will be derived later, between data $(A, B, C_j)$ and data $(a(\xi), B(\xi), c_j)$ for time $T \geq 0$ and for time $t \geq 0$ we then come to known relations for coordinates $(x,t)$.

Let us rewrite the linear problem (9) and the time development, for time $t$, (12) into coordinates $(X,T)$:

$$\frac{\partial v_1}{\partial X} = (-\frac{i.\zeta}{2} + \frac{i}{8.\zeta}\cos(u))v_1 + (\frac{i}{8\zeta}\sin(u) - \frac{1}{4}(\frac{\partial u}{\partial X} + \frac{\partial u}{\partial T}))v_2 \quad (79)$$

$$\frac{\partial v_2}{\partial X} = (\frac{1}{4}(\frac{\partial u}{\partial X} + \frac{\partial u}{\partial T}) + \frac{i}{8.\zeta}\sin(u))v_1 + (\frac{i.\zeta}{2} - \frac{i}{8.\zeta}\cos(u))v_2$$

and

$$\frac{\partial v_1}{\partial T} = (-\frac{i.\zeta}{2} - \frac{i}{8.\zeta}\cos(u))v_1 - (\frac{i}{8\zeta}\sin(u) + \frac{1}{4}(\frac{\partial u}{\partial X} + \frac{\partial u}{\partial T}))v_2 \quad (80)$$

$$\frac{\partial v_2}{\partial T} = (\frac{1}{4}(\frac{\partial u}{\partial X} + \frac{\partial u}{\partial T}) - \frac{i}{8.\zeta}\sin(u))v_1 + (\frac{i.\zeta}{2} + \frac{i}{8.\zeta}\cos(u))v_2.$$

For the eigenvalue $\zeta$ it holds that $\frac{\partial \zeta}{\partial t} = 0$ and that $\frac{\partial \zeta}{\partial x} = 0$, and from this it follows that:

$$\frac{\partial \zeta}{\partial T} = 0, \quad (81)$$

$$\frac{\partial \zeta}{\partial X} = 0.$$

Jost functions $\Phi$ and $\Psi$ are such solutions of (77) which fulfill:

$$\Phi \longrightarrow \begin{pmatrix} 1 \\ 0 \end{pmatrix} . \exp(-\frac{i}{2}.(\zeta - \frac{1}{4.\zeta}).X - \frac{i}{2}.(\zeta + \frac{1}{4.\zeta}).T) \quad (82)$$

for

$$X \longrightarrow -\infty$$

and

$$Im\zeta \geq 0,$$

and

$$\Psi \longrightarrow \begin{pmatrix} 0 \\ 1 \end{pmatrix} . \exp(+\frac{i}{2}.(\zeta - \frac{1}{4.\zeta}).X + \frac{i}{2}.(\zeta + \frac{1}{4.\zeta}).T) \quad (83)$$

for

$$X \longrightarrow +\infty$$

and

$$Im\zeta \geq 0.$$

Let us define again:

$$\overline{\Psi}(X, \zeta) \equiv \begin{pmatrix} \Psi_2^*(X, \zeta^*) \\ -\Psi_1^*(X, \zeta^*) \end{pmatrix} \quad (84)$$

Because $\Psi$ and $\overline{\Psi}$ are linearly independent for $\zeta = \zeta^* = \xi$, we can again write:

$$\Phi(X, \xi) = A_0 \overline{\Psi(X, \xi)} + B_0(\xi)\Psi(X, \xi). \quad (85)$$

Here $A_0(\xi)$ can be analytically extended into upper halfplane $Im\zeta > 0$. Zeros in $A_0(\xi)$ are discrete eigenvalues (77) and this is the reason that they are eigenvalues of the initial problem (22), we have $\zeta_j$ where $j = 1, 2, 3, ..., N$. For $\zeta = \zeta_j$ we have:

$$\Phi(X, \zeta_j) = C_{j0}.\Psi(X, \zeta_j). \quad (86)$$

Coefficients $A_0(\xi)$, $B_0(\xi)$ and $C_{j0}$ are independent on time $T$ (in difference from $b(\xi)$ and from $c_j$, which are dependent on time $t$, this is the reason why the index "0" occurs in coefficients $A$, $B$ and $C$ !). We will easily show from (78) for $X \longrightarrow +\infty$ that $\mathbf{v} = \overline{\Psi} + \frac{B}{A}.\Psi$ and that for $X \longrightarrow -\infty$ we have $\mathbf{v} = \frac{1}{A}.\Phi$. By solving the direct problem of the Theory of Scattering for (77) we find from the initial condition $u(X, T = 0)$ coefficients $A_0(\xi)$, $B_0(\xi)$ and $C_{0j}$ for $j = 1, 2, 3, ..., N$. These coefficients are in relation to the coefficients $a(\xi)$, $b(\xi)$ and $c_j$ for $j = 1, 2, 3, ..., N$ :

$$a(\xi, t = 0) = a_0(\xi) = A_0(\xi), \quad (87)$$

$$b(\xi, t = 0) = b_0(\xi) = B_0(\xi),$$

$$c(\xi, t = 0) = c_{0j} = C_{0j},$$

where $j = 1, 2, 3, ..., N$.

# 10 Solution of the Sine-Gordon Equation by the Inverse Method of Scattering.

We would like to solve the Sine-Gordon Equation (1):

$$\frac{\partial^2}{\partial x \partial t} u(x,t) = \sin(u(x,t)) \tag{88}$$

with the initial condition $u(x, t = 0)$ and the boundary condition

$$\lim_{x \longrightarrow \pm\infty} u(x,t) = 0.$$

This limit is with ( mod $2.\pi$). Moreover we assume $\lim_{x \longrightarrow \pm\infty} \frac{\partial^n x}{\partial x^n} u(x,t) \longrightarrow 0$ exponentially.

By using the direct problem of the Theory of Scattering we find for the linear problem (7), where $q = -\frac{1}{2}\frac{\partial}{\partial x} u(x, t = 0)$, the scattering data at time $t = 0$:

$$a(\xi, t = 0), b(\xi, t = 0), \tag{89}$$

$$\zeta_j$$

where $j = 1, 2, 3, ..., N$, and

$$c_{0j}.$$

Knowing these data we find with using time development of parameters of scattering (see the Section: Time Development of the Scattering Data.) scattering data for any time $t > 0$:

$$a(\xi, t) = a(\xi, t = 0), \tag{90}$$

$$b(\xi, t) = b(\xi, t = 0). \exp(-\frac{i.t}{2.\xi}),$$

$$c_j = c_j(t = 0). \exp(-\frac{i.t}{2.\zeta_j}),$$

$$\zeta_j(t = 0) = \zeta(t),$$

where $j = 1, 2, 3, ..., N$. From the scattering data in time $t$ we reconstruct the potential $q(x,t)$ using the Method of Inverse problem of the Scattering Theory. Due to the fact that:

$$q(x,t) = -\frac{\partial}{\partial x} u(x,t), \tag{91}$$

we obtain from this equation the solution of the Sine-Gordon Equation (1). Fact that we obtain such a solution follows from the formulation of linear problem associated with the Sine-Gordon Equation.

We can describe steps of the method: 1, we have Sine-Gordon Equation and initial conditions at $t = 0$, 2, by the Direct Method of the Theory of Scattering we obtain 3, scattering data at time $t = 0$ 4, time development of scattering data leads to 5, scattering data at time $t > 0$, from these data by the inverse problem of the Theory of Scattering we "reconstruct" the potential $q(x,t)$ from which we obtain 6, the solution of the Sine-Gordon Equation $u(x,t)$ from the step 1, .

# 11 Soliton Solutions - Kinks.

At practical calculations searching for soliton solutions of the Sine-Gordon Equation (89) we start with choosing the scattering data in time $t = 0$. It is due to the fact that there is the following correspondence between number of discrete eigenvalues, their concrete form and the character of the soliton solution of the Sine-Gordon Equation:

- To one purely imaginary eigenvalue $\zeta = i.\eta$, where $\eta > 0$, the kink solution of the type (4) corresponds to:

$$u(X,T) = 4.\arctan(\exp(\pm\theta))$$

  where

$$\theta \equiv (\eta + \frac{1}{4.\eta})(X - X_0) + (\eta - \frac{1}{4.\eta}).T$$

- To N-different purely imaginary $\zeta_j$, where $j = 1, 2, 3, ..., N$ and where $\zeta_j > 0$ for every $j$, and ($\eta_i \neq \eta_j$) for $i \neq j$ corresponds to N-soliton solution (6).

- To a pair of complex eigenvalues (not purely imaginary) corresponds a solution of the type breather (5): $\zeta$, $-\zeta^*$ (where $|\zeta| = \frac{1}{2}$ in this special case).

- To a general initial condition $u(x, t = 0)$ it will corresponds a solution which will be obtained by mixing L kinks, M breathers and some oscillating field.

## 11.1 Kink.

In this case one bounded state in the linear problem (7) is characterized by one discrete value $\zeta = i.\eta$ where $\eta > 0$ at the time $t = 0$. We will assume that we have a reflexless potential, e.i.

$$b(\xi, t = 0) = 0.$$

The coefficient $a(\xi)$ remains not specified, we know that after analytical continuation into upper halfplane $Im\zeta > 0$ this coefficient has in $\zeta = i.\eta$ zero, $a(i.\eta) = 0$. Time development of the scattering data is:

$$a(\xi, t) = a(\xi, t = 0),$$

$$\zeta(t) = \zeta(t = 0) = i.\eta,$$

$$c_1(t) = c_1(t = 0).\exp(-\frac{t}{2.\eta}),$$

$$N = 1,$$

here $\eta > 0$. Constants $a(\xi, t = 0)$ and $c_1(t = 0)$ are occurring in the system of equations of the inverse problem of the Theory of Scattering (62), (63), (64) and (65) only in the form of ratio: $\widetilde{c}_1 = \frac{c_1(t)}{\frac{da(\zeta)}{d\zeta}|_{\zeta=i.\eta}} = \widetilde{c}_1(t=0).\exp(-\frac{t}{2.\eta})$. The value of $\widetilde{c}_1(t)$, e.i. in equations (62), (63), (64) and (65) $c_1$ without tilde, in the time $t = 0$ will be chosen later.

From equations (62) and (63) we obtain:
$$\Theta_1(\xi, x) = 0,$$
$$\Theta_2(\xi, x) = 0,$$
because $b = 0$ and according to (61) we have $c(\xi, x) = 0$.

Equations (64) and (65) are taking the form (here $t > 0$ and $x$ are fixed):
$$\psi_1(x, i.\eta).\exp(\eta.x) + \frac{\exp(-\eta.x)}{i.\eta - (-i.\eta)}.c_1^*.\psi_2^*(x, i.\eta) = 0$$

$$-\frac{\exp(-\eta.x)}{(-i.\eta) - i.\eta}.c_1\psi_1(x, i.\eta) + \psi_2^*.\exp(+\eta.x) = 1$$

from which we obtain:
$$\psi_2^*(x, i.\eta) = \frac{1}{\exp(+\eta.x) + \frac{\exp(-3.\eta.x)}{4.\eta^2}\mid c \mid^2} \tag{92}$$

and
$$\psi_1(x, i.\eta) = \frac{-c^*.\exp(-2.\eta.x)}{2.i.\eta.(\exp(+\eta.x) + \frac{\exp(-3.\eta.x)}{4.\eta^2}\mid c \mid^2)}. \tag{93}$$

Let us substitute (92) into (68). We obtain:
$$q(x, t) = -2.i.c_1^*.\exp(-\eta.x)\frac{1}{\exp(+\eta.x) + \frac{|c|^2}{4.\eta}.\exp(-3.\eta.x)} =$$

$$-\frac{2.i.c_0^*}{\mid c_0 \mid}.\eta.\frac{1}{\cosh(2.\eta.x + \frac{t}{\eta} + \delta)},$$

where we put $\exp(-\delta) \equiv \frac{|c_0|}{2.\eta} > 0$. Let us choose $c_0 = i.\gamma$ in order the potential $q$ was real, as it should be for the Sine-Gordon Equation, and we obtain:
$$q(x, t) = -2.\eta.\frac{1}{\cosh(2.\eta.x + \frac{t}{\eta} + \delta)}, \tag{94}$$

here $\exp(-\delta) = \frac{|\gamma|}{2.\eta}$. Using the relation $q(x,t) = -\frac{1}{2}\frac{\partial}{\partial x}u(x,t)$ we obtain the function $u(x,t)$:
$$u(x, t) = 4.\arctan(\exp(+2.\eta.x + \frac{t}{\eta} + \delta)). \tag{95}$$

After transition to the $(X, T)$ coordinates we obtain solution $u(X, T)$ from the beginning of this Section with the condition for the velocity $U$ of the movement of soliton:
$$-1 \leq U = \frac{\eta - \frac{1}{4.\eta}}{\eta + \frac{1}{4.\eta}} \leq 1. \tag{96}$$

To soliton which has zero velocity corresponds $\eta = \frac{1}{2}$. The velocity $U \longrightarrow -1$ corresponds with $\eta \longrightarrow 0_+$. The velocity $U \longrightarrow +1$ corresponds with $\eta \longrightarrow +\infty$. The "relativistic" factor $\sqrt{1 - U^2}$ (with "velocity of light" $= 1$) corresponds to the Lorentz symmetry of the Sine-Gordon Equation.

Let us note that to the soliton (kink) solution (95) the corresponding initial condition is:
$$u(x, t = 0) = 4.\arctan(\exp(2.\eta.x + \delta)) \tag{97}$$
for the Sine-Gordon Equation $\frac{\partial^2}{\partial x \partial t}u(x, t) = \sin(u(x, t))$.

# 12   Discussion.

This introductory paper may be called an opening paper to our aim to study some problems of nonlinear dynamics. It introduces Inverse Scattering Method, the Sine-Gordon Equation is described as an example (well known).

Our first region of interest is to study solitary traveling wave solutions of pressure equation for bubbly liquids with examination for viscosity and heat transfer [20]. Authors in their research succeeded in applying three different methods on one of the most important model in fluid mechanics which is pressure equation of bubbly liquids with examination for viscosity and heat transfer, e.i. the Kudryashov-Sinelshchikov Equation. This equation describes the pressure waves in the liquid with gas bubbles taking into account the heat transfer and viscosity. Authors get many forms of exact and solitary traveling wave solutions. They present a good comparison between the methods and experimental results. Kudryashov-Sinelshchikov Equation [21] gives a description of the pressure waves in a mixture liquid and gas bubbles taking into consideration the viscosity of liquid and the heat transfer and dynamics of contrast agents, f.e. in the blood flow at ultrasonic researches.

Our second region of interest is to study motions in transport, e.i motions of automobiles, trains and ships. The nonlinear dynamics of ship motions was recently overviewed in [22] including some recent developments. In the field of nonlinear ship dynamics we explore the onset and evolution of the unfamiliar, and often unsafe, dynamic responses beyond the linearity regime, which are not amenable to the conventional techniques of seakeeping theory, as authors writes. The presence of nonlinearity in the relation between excitation and response in a dynamical system creates the prospect of having multiple solutions for certain values of the system´s parameters. This could further lead to a very complicated phenomena even when the type of nonlinearity is very simple; such as a quadratic or cubic term in the stiffness component of an otherwise ordinary linear driven oscillator. The above ideas are not yet fully domesticated in transport architecture.